# On a characterization of cellular automata in tilings of the hyperbolic plane

Maurice Margenstern[1]

Laboratoire d'Informatique Théorique et Appliquée, EA 3097,
Université de Metz, I.U.T. de Metz,
Département d'Informatique,
Île du Saulcy,
57045 Metz Cedex, France,
`margens@univ-metz.fr`

**Abstract.** In this paper, we look at the extention of Hedlund's characterization of cellular automata to the case of cellular automata in the hyperbolic plane. This requires an additionnal condition. The new theorem is proved with full details in the case of the pentagrid and in the case of the ternary heptagrid and enough indications to show that it holds also on the grids $\{p,q\}$ of the hyperbolic plane.

## 1  Introduction

Hedlund's theorem, see [2] is a well known characterization of cellular automata in terms of transformation over the space of all possible configurations. The theorem says that the global transition function defined by the local rule of a cellular automaton is a continuous function on the space of all configurations of the cellular automaton and that this global function also commutes with all shifts. The theorem states that the converse is true. As a well known corollary of the theorem, we know that a cellular automaton is reversible if and only if its global transition function is bijective.

In the paper, we investigate the status of the theorem in the case of cellular automata in the hyperbolic plane. We shall prove that it is not true, *stricto-sensu*: there are cellular automata in the hyperbolic plane which do not commute with all the shifts which leave invariant the grid of the cellular automaton. In fact, we shall prove that the commutation with shifts entails another property of the cellular automaton which we call **rotation invariance**. Then, denoting $\mathcal{C}$ the space of configurations for the considered grid, here the pentagrid or the ternary heptagrid. We can state:

**Theorem 1** *A mapping $F$ from $\mathcal{C}$ into $\mathcal{C}$ is the global transition function of a rotation invariant cellular automaton on the pentagrid or the ternary heptagrid if and only if $F$ is continous and if $F$ commutes with all the shifts leaving the grid invariant.*

Later, we shall extend the theorem to all grids of the form $\{p,q\}$ of the hyperbolic plane. During the proof, we shall prove that the considered shifts are finitely generated: in the case of the pentagrid and of the ternary heptagrid but also, generally, for any grid $\{p,q\}$.

As we shall see, the main concern of the proof is the coordinate system for locating the cells of the cellular automaton.

This problem is obvious in the case of the Euclidean plane: in fact, whatever the grid, we may consider that we are in $\mathbb{Z}^2$ and the proof is almost word by word the same as in the unidimensional case.

In the case of the hyperbolic plane, things are very different. First, there are infinitely many tilings defined by tessellation, *i.e.* generated by the reflection of a regular polygon in its edges and, recursively, of the images in their edges. Second, there is no as general pattern as in the Euclidean plane to locate the cells of the grid. In [4], a new tool was introduced which allows to better handle the problem. It gives a general frame to locate the cells in any grid $\{p,q\}$, but the realization of the frame for each tiling $\{p,q\}$ generally depends of the tiling. There are a few exceptions. Among them we have the case of the pentagrid and of the ternary heptagrid which, up to a point, can be handled in the same way.

Just after this introduction, in the second section, we remind the reader with the system of coordinates introduced in [4], also explained in [5]. Then, in the third section, we look at the continuity part of the theorem. In the fourth section, we prove that the shifts are finitely generated, extending the result to any grid $\{p,q\}$. In the fifth section, we prove that the commutation with the shifts is equivalent to the rotation invariance. In the sixth section, we prove the theorem and its corollary about reversible cellular automata in the hyperbolic plane.

The reader is referred to [5] for an introduction to hyperbolic geometry which is aimed at the implementation of cellular automata in the corresponding spaces.

## 2 Coordinates in the pentagrid and in the heptagrid of the hyperbolic plane

As recalled in the introduction, the hyperbolic plane admits infinitely many tilings defined by tessellation. This is a corollary of a famous theorem proved by Henri POINCARÉ in the late 19[th] century, see [5], for instance.

Figure 1 sketchily remembers that the tiling is spanned by a generating tree. Now, as indicated in figure 2, five quarters around a central tile allows us to exactly cover the hyperbolic plane with the **pentagrid** which is the tessellation obtained from the regular pentagon with right angles.

In the right-hand side picture of figure 2, we remember the basic process which defines the coordinates in a quarter of the pentagrid, see [5]. We number the nodes of the tree, starting from the root and going on, level by level and, on each level, from the left to the right. Then, we represent each number in the basis defined by the Fibonacci sequence with $f_1=1$, $f_2=2$, taking the maximal representation, see[4,5].

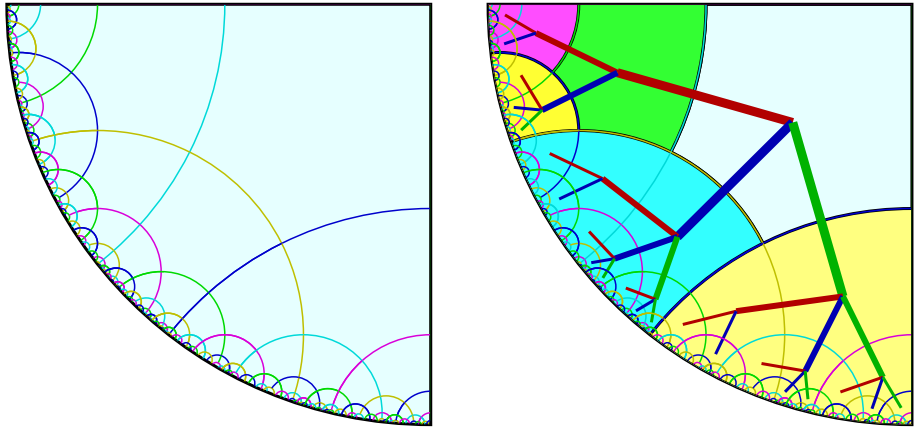

**Figure 1** *On the left: the tiling; on the right: the underlying tree which spans the tiling.*

From the left-hand side picture of figure 2, we can see that any tile can be located by the indication of two numbers $(i, \nu)$, where $i \in \{1..5\}$ numbers a quarter around the central tile and $\nu$ is the number of the tile in the corresponding tree which we call a **Fibonacci tree** as the number of tiles at distance $n$ from the root of the tree is $f_{2n+1}$, see [6,4,5].

Almost the same system of coordinates can be defined for the **ternary heptagrid** which is obtained by tessellation from a regular heptagon with the interior angle of $\dfrac{2\pi}{3}$, see figure 3.

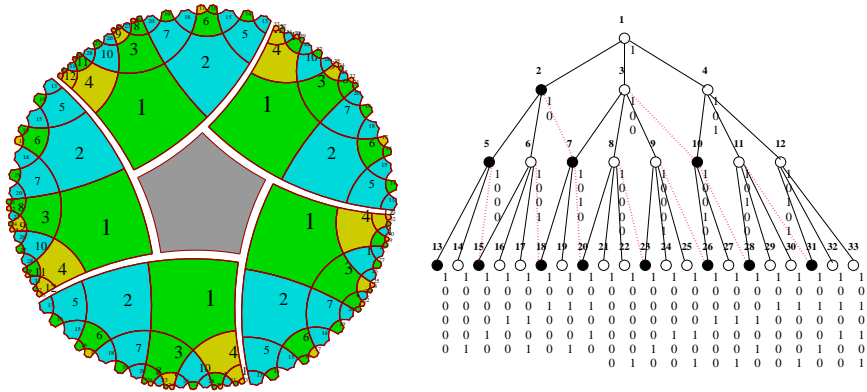

**Figure 2** *On the left: five quarters around a central tile; on the right: the representations of the numbers attached to the nodes of the Fibonacci tree.*

Remind that the main reason of this system of coordinates is that from any cell, we can find out the coordinates of its neighbours in linear time with respect to the coordinate of the cell. Also in linear time from the coordinate of the cell, we can compute the path which goes from the central cell to the cell.

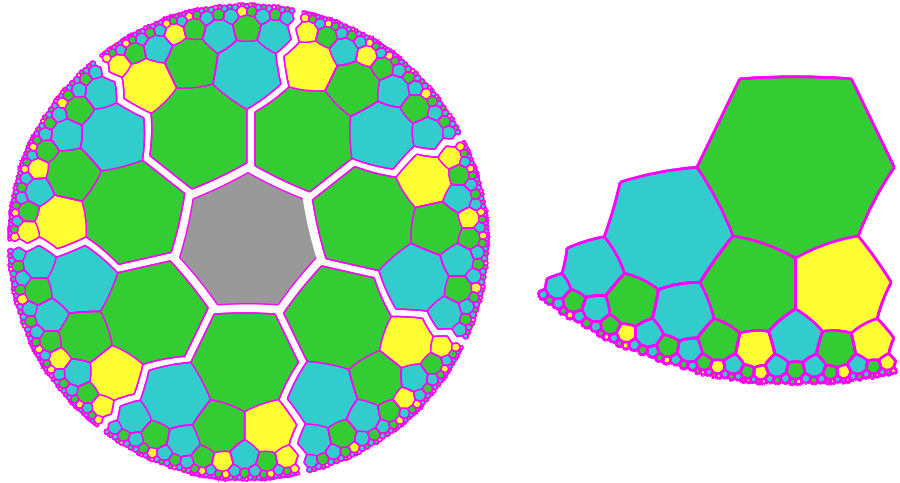

**Figure 3** *On the left: seven sectors around a central tile; on the right: the structure of a sector, where a Fibonacci tree can easily be recognized.*

Now, as the system coordinate is fixed, we can turn to the space of configurations.

## 3   Topology on the space of all possible configurations

In the proof of Hedlund's theorem, the space of configurations a cellular automaton with $Q$ as a set of states is represented by $Q^{\mathbb{Z}^2}$. Accordingly, each configuration is viewed as a mapping from $\mathbb{Z}^2$ into $Q$. Now, as $Q$ is a finite set, it is naturally endowed with the discrete topolgy which can be defined by a distance: $\text{dist}(q_1, q_2) = 1$ if $q_1 \neq q_2$ and $\text{dist}(q_1, q_2) = 0$ if $q_1 = q_2$. The space $Q^{\mathbb{Z}^2}$ is endowed with the product topology. It is the topology of the simple convergence, and it can also be defined by a distance:

$$\text{dist}(x, y) = \sum_{i \in \mathbb{Z}^2} \frac{\text{dist}(x(i), y(i))}{4(2|i|+1)} 2^{-|i|},$$

where $|(\alpha, \beta)| = \max(|\alpha|, |\beta|)$. Note that $4(2n+1)$ is the length of a square centred at (0,0), exactly containing the points $(\alpha, \beta)$ with $|(\alpha, \beta)| = n$.

The translation to the case of the pentagrid or the heptagrid is immediate. Again, let $Q$ be the set of states of the cellular automaton. We define dist on $Q$ as previously. Now, we denote by $\mathcal{F}_5$ the set of five Fibonacci trees dispatched around a central node. Similarly, we define $\mathcal{F}_7$ for the set of seven Fibonacci trees dispatched in a similar way.

Then the distance on the set of all configurations is defined by
$$\mathrm{dist}(x,y) = \sum_{i \in \mathcal{F}_\alpha} \frac{\mathrm{dist}(x(i),y(i))}{\alpha(f_{2|i|+1})} 2^{-|i|},$$
where $\alpha \in \{5,7\}$ and $|i|$ is defined by the distance of $i$ to the central cell. In other terms, $|i|$ is the index of the level of the tree on which $i$ is. We note that $\alpha f_{2n+1}$ is the number of nodes which are at distance $n$ from the central cell.

It is not difficult to see that if $x(i) = y(i)$ on a ball of radius $n$ around the central cell, $\mathrm{dist}(x,y) \leq 2^{-n}$. Conversely, if $\mathrm{dist}(x,y) \leq \dfrac{1}{f_{2n+1}2^{-n}}$, then $x(i) = y(i)$ on a ball of radius $n-1$ around the central cell.

As well known, the set of all configurations $Q^{\mathcal{F}_\alpha}$ endowed with the just defined topology is a compact metric space.

It is plain that we have the following property:

**Lemma 1** *A cellular automaton on the pentagrid or on the heptagrid is continuous on the set of all configurations with respect to the product topology.*

Indeed, as long as two configurations are equal on the neighbourhood of a cell $c$ which corresponds to the local function of transition, the values given by the cellular automaton at $c$ are the same for both configurations.

It is possible to extend this result to any grid $\{p,q\}$.

Remind that the restriction of the tiling to an angular sector of angle $\dfrac{2\pi}{q}$ can be spanned by a tree $\mathcal{F}_{pq}$, see [7]. Accordingly, the whole tiling can be generated by $p.(h{-}1)$ trees dispatched around a central tile, where $h = \lfloor \dfrac{q}{2} \rfloor$. Then, there is a bijection between the copies of the spanning trees and this tile with the tiling. Let $\mathscr{F}_{pq}$ denote the new tree obtained by the central cell surrounded by the $p.(h{-}1)$ copies of $\mathcal{F}_{pq}$. We can then consider that the set of configurations of a cellular automaton $A$ in the grid $\{p,q\}$ is $Q^{\mathscr{F}_{pq}}$, where $Q$ is the set of states of $A$.

Then, the metric of this compact metric space is defined by:
$$\mathrm{dist}(x,y) = \sum_{i \in \mathscr{F}_{pq}} \frac{\mathrm{dist}(x(i),y(i))}{\alpha(u_i)} 2^{-|i|},$$
where $u_i$ is the number of nodes at distance $i$ from the root of $\mathcal{F}_{pq}$, and where $\alpha = p(h{-}1)$, as there are $p(h{-}1)$ copies of $\mathcal{F}_{pq}$ around the considered central cell. Note that the case $q = 3$ has an exceptional status, see [5].

Now, the same arguments as above for the pentagrid and for the ternary heptagrid allows us to reformulate lemma 1 as:

**Lemma 2** *For all positive integers $p$ and $q$ with $\dfrac{1}{p} + \dfrac{1}{q} < \dfrac{1}{2}$, a cellular automaton on the grid $\{p,q\}$ of the hyperbolic plane is continuous on the set of all configurations with respect to the product topology.*

## 4   Generating the shifts

First, if we analyze the proof of Hedlund's theorem, we only need the commutation with shifts to prove that a continous mapping on the set of configurations is a cellular automaton. It is not required that the shifts constitute a group. What is needed is that for any cell $c$, there is a shift which transforms the origin $(0,0)$ into $c$. Next, if the shifts we need can be generated by finitely many fixed in advance shifts, we are done, whether the shifts commute or not between themselves. If they don't commute, the representation will be more complicate, but this aspect is not relevent for our question.

The second good news is that we can find two shifts for the generation of all the shifts, both in the case of the pentagrid and of the ternary heptagrid. The proof is rather simple for the pentagrid. It is a bit more complex for the ternary heptagrid. It is a bit more difficult, also in the case of the grids $\{p,q\}$, when $q$ is even. At last, it requires some effort in the case of the grids $\{p,q\}$, when $q$ is odd.

In all these studies, we shall make use of the following general property:

**Lemma 3** *Let $\tau_1$ and $\tau_2$ be two shifts along the lines $\ell_1$ and $\ell_2$ respectively. Then, $\tau_1 \circ \tau_2 \circ \tau_1^{-1}$ is a shift along the line $\tau_1(\ell_2)$, with the same amplitude as $\tau_2$ and in the same direction.*

Although it is well know in the specialized literature, we provide the reader with a proof of the lemma. It relies on the following well known features on shifts in the hyperbolic plane:

  $(i)$ a shift has no fixed point in the hyperbolic plane,
  $(ii)$ there is a unique line of the hyperbolic plane, called the **axis** of the shift which is globally invariant under the action of the shift,
  $(iii)$ a shift is an isometry, in particular it preserves lengths and it transforms lines into lines

A transformation of the hyperbolic plane into itself which satisfies these three properties is a shift along its axis.

Proof of lemma 3. Consider two shifts $\tau_1$ and $\tau_2$, and let $\tau = \tau_1 \circ \tau_2 \circ \tau_1^{-1}$. Let $\delta$ be the axis of $\tau_2$ and let $\delta_1 = \tau_1(\delta)$. Take a point $A$ on the line $\delta$ and define $A_1 = \tau_1^{-1}(A)$. Clearly, if $\tau_2(A) = B$, we have $\tau(A_1) = \tau_1(B)$. Define $B_1 = \tau_1(B)$. Now, as $\delta$ is the axis of $\tau_2$, $B \in \delta$ and so, $A_1, B_1 \in \delta_1$. Now, $\tau_1(B_1) = B$, so that $\tau(B_1) = \tau_1(C)$, where $C = \tau_2(B)$. As $\delta$ is the axis of $\tau_2$ and as $B \in \delta$, we have also that $C \in \delta$, so that $\tau_1(C) \in \delta_1$. Now, $\tau_1(C) = \tau(B_1)$, so that $\tau(B_1) \in \delta_1$. Accordingly, $A_1$ and $B_1$ belong to $\delta_1$ and $A_1 \neq B_1$ as $A \neq B = \tau_2(B)$ as $\tau_2$ has no fixed point. Consequently, as $\tau$ is an isometry as a finite product of isometries, $\tau(\delta_1) \subseteq \delta_1$. And so, $\delta_1$ is the axis of $\tau$. Also, $\tau$ has no fixed point. Indeed, if $P$ were a fixed point of $\tau$, $\tau_1^{-1}(P)$ would be a fixed point of $\tau_2$. Impossible, as $\tau_2$ is a shift. Accordingly, $\tau$ is a shift along $\delta_1$. Now, $A_1 B_1 = \tau_1^{-1}(AB) = AB$, as $\tau_1$ is an isometry. And so the amplitude of $\tau$, which is $A_1 B_1 = A_1 \tau(A_1)$, is $AB = A\tau_2(A)$, the amplitude of $\tau_2$. ∎

Now, it is possible to state:

**Lemma 4** *The shifts leaving the pentagrid globally invariant are generated by two shifts and their inverses. The same property holds for the ternary heptagrid.*

We shall consider the cases of the pentagrid and of the heptagrid separately. We shall make use of the traditional notation of $\tau_1 \circ \tau_2 \circ \tau_1^{-1}$ by $\tau_2^{\tau_1}$.

First, consider the case of the pentagrid, it is illustrated by the left-hand side picture of figure 4.

Fix a tile of the pentagrid, say $\Pi_0$. Fix an edge of $\Pi_0$ and let $\ell_1$ be the line which supports this edge. Consider a contiguous edge, supported by the line $\ell_2$. Both lines are lines of the pentagrid. Let $A$ be the common point of $\ell_1$ and $\ell_2$: it is a vertex of $\Pi_0$. Let $B$ be the other vertex of $\Pi_0$ on $\ell_1$ and let $C$ be the other vertex of $\Pi_0$ on $\ell_2$. Then, define $\tau_1$ to be the shift along $\ell_1$ which transforms $A$ into $B$ and define $\tau_2$ to be the shift along $\ell_2$ which transforms $A$ into $C$. Now, let us show that $\tau_1$, $\tau_2$, $\tau_1^{-1}$ and $\tau_2^{-1}$ generate all the shifts which leave the pentagrid globally invariant. It will be enough to show that if we take a tile $P$, there is a product of $\tau_1$, $\tau_2$, $\tau_1^{-1}$ and $\tau_2^{-1}$ which is a shift and which transforms $\Pi_0$ into $P$.

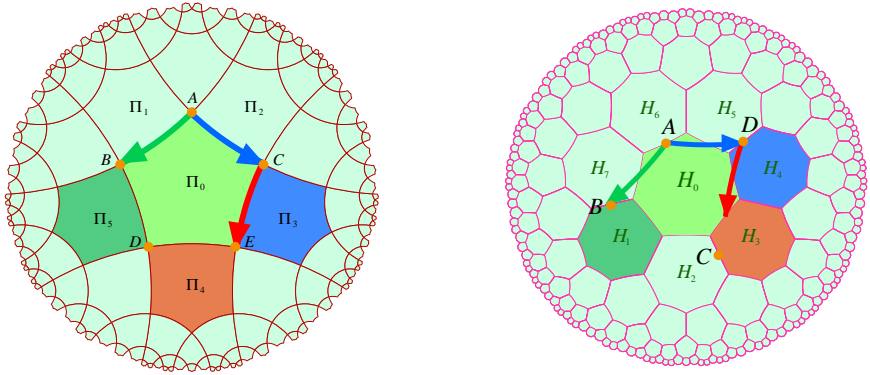

**Figure 4** *Action on the shifts $\tau_1$, green, and $\tau_2$, blue. On the left-hand side, $\Pi_i$, $i \in \{1..5\}$ denote the neighbour of $\Pi_0$ sharing with it the edge $i$. Similarly, on the right-hand side, the neighbours of $H_0$ are denoted by $H_i$, $i \in \{1..7\}$. Also, note the mid-points $A$, $B$, $C$ and $D$ which are used by table 1.*

Number the edges of $\Pi_0$ by 1 up to 5 and assume that the edge 1 is $AB$ and that the edge 2 is $AC$. Then, from lemma 3, $\tau_2^{\tau_1}$ is a shift along the edge 5, transforming $B$ into the other end of this edge. Similarly, $\tau_1^{\tau_2}$ is the shift along the edge 3 which transforms $C$ into the other end of this edge. Now, it is not difficult to see that $\tau_2^{\tau_1^{\tau_2}}$ is a shift along the edge 4 transforming $\tau_2^{\tau_1}(B)$ into $\tau_1^{\tau_2}(C)$. Taking these shifts and the inverses, we get shifts which transform $\Pi_0$ in all its neighbouring tiles in the sense of Moore. Now, it is not difficult to repeat this construction with any neighbour of $\Pi_0$: it shares an edge with $\Pi_0$ and it has two other edges which are supported by a line which also supports another edge of $\Pi_0$. Accordingly, we get all the tiles within a ball of radius 2 around $\Pi_0$.

Now, by an easy induction, we get all the tiles of the pentagrid. Note, that for a given shift of the pentagrid, there is no unique representation of this shift as a product of powers of $\tau_1$, $\tau_2$ and their inverses.

Now, let us look at the case of the ternary heptagrid which is illustrated by the right-hand side picture of figure 4.

This time, we cannot take the lines which support the edges of a heptagon: due to the angle $\frac{2\pi}{3}$, such a line supports edges but it also cuts heptagons for which they are an axis of reflection. In [1,5], I have indicated that **mid-point lines** play the rôle of the expected shifts. This is what is performed in the right-hand side picture of figure 4.

Consider again $\tau_2^{\tau_1}$, where $\tau_1$ and $\tau_2$ are shifts along two mid-point lines which meet on an edge of the heptagon. By construction of the mid-point lines, the definition of $\tau_1$ and $\tau_2$ involve the neighbours of $H_0$, the heptagon which we fix in order to define the generators of the shifts. As shown in the right-hand side of figure 4, $\tau_1$ transforms $H_0$, say into $H_1$ while $\tau_2$ transforms $H_0$ into $H_4$: we number the edges of $H_0$ clockwise. Now $\tau_2^{\tau_1}$ transforms $H_1$ into $H_2$, and so, it transforms $H_0$ into $H_3$. Similarly, we find that $\tau_1^{\tau_2}$ transforms $H_0$ into $H_2$.

For the convenience of the reader, we indicate the next shifts which transform $H_0$ into the remaining neighbours. Using the previous transformations, let us set $\xi_1 = (\tau_2^{\tau_1})^{-1}$ and $\xi_2 = (\tau_1^{\tau_2})^{-1}$. Then, $\xi_1$ transforms $H_0$ into $H_7$ while $\xi_2$ transforms $H_0$ into $H_5$. At last, $\xi_1^{\xi_2}$ transforms $H_0$ into $H_5$.

| $H_i$ | point | shift$_1$ | shift$_2$ |
|---|---|---|---|
| $H_1$ | $B$ | $\tau_1$ | $\tau_2^{\tau_1}$ |
| $H_2$ | $C$ | $\tau_1^{\tau_2}$ | $\xi_1$ |
| $H_3$ | $C$ | $\xi_1$ | $\xi_2$ |
| $H_4$ | $D$ | $\xi_2$ | $\tau_2$ |
| $H_5$ | $D$ | $\tau_2^{-1}$ | $\xi_2$ |
| $H_6$ | $B$ | $\tau_1^{-1}$ | $\tau_2^{-1}$ |
| $H_7$ | $B$ | $\tau_1^{-1}$ | $\xi_1$ |

**Table 1** *The shifts which, for each neighbour of $H_0$ generate the transformations of $H_i$ into its neighbours. Note that there is no order in the pair of generating shifts.*

In order to reproduce the same actions for the neighbours of $H_0$, we just need to define mid-points of edges which will allow us to define the shifts which will play the rôle of $\tau_1$ and $\tau_2$ for each neighbour. The considered mid-points are indicated in the right-hand side picture of figure 4. Table 1 indicates for each neighbour the mid-point which is used and the shifts denoted in terms of the shifts which we already defined.

This allows us to prove the statement of lemma 4 for the ternary heptagrid. ∎

Before proving the same property of finite generation for any grid $\{p,q\}$ of the hyperbolic plane, the reader may wonder why we need two different techniques for the pentagrid and for the heptagrid? The mid-point lines can also be defined in the pentagrid and the same type of shifts defined for the ternary heptagrid can be defined for the pentagrid. This is true but such shifts would not be interesting for our purpose in the pentagrid. In the pentagrid, it is possible to colour the tiles with black and white in order to get something which looks like a chessboard: any white tile is surrounded by black ones and any black one is surrounded by white ones. Now, it is not difficult to remark that the shifts based on mid-point lines transform a tile of one colour into a tile of the same colour. Accordingly, we cannot get the immediate neighbours of a cell with such shifts.

As announced in our introduction, now, we prove the same property of finite generation for any grid $\{p,q\}$ of the hyperbolic plane. From the previous remark, we may expect that the parity of $q$ is important.

Indeed, the argument which we considered can be extended to any grid $\{p,q\}$ but, roughly speaking, the argument for the pentagrid extends to all grid $\{p,q\}$, when $q$ is even. Similarly, the argument for the ternary heptagrid extends to all grid $\{p,q\}$, when $q$ is odd.

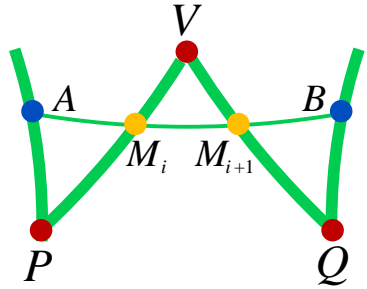

**Figure 5** *The mid-point figure around a vertex, when $q$ is odd.*

This is obvious for the grids $\{p,4\}$ and $\{p,3\}$. For the other grids, it follows from the following consideration. When $q$ is bigger, number the $p$ edges of the basic polygon $P_0$, $e_1,\ldots,e_p$, by turning around $P_0$, clockwise. Also number the vertices $V_1,\ldots,V_p$ with $V_{i+1} \in e_i \cap e_{i+1}$ for $i \in \{1..p-1\}$ and $V_1 \in e_1 \cap e_p$. Denote by $\tau_i$ the shift along the axis of $e_i$ which transforms $V_i$ into $V_{i+1}$, considering that $V_{p+1} = V_1$. Then, if we perform successively the shifts $\tau_1,\ldots,e_p$, the image of $e_1$ is not $e_1$ but its image under a rotation of $p.\dfrac{2\pi}{q}$ around $V_1$. Repeating this *tour*, we get all the tiles which are around $V_1$. Now, from $\tau_1$, we go to a polygon $P$ which is around $V_2$. With an appropriate number of rounds around $P$, we get the neighbour of $P_0$ which shares $e_2$ with it. And then, we can repeat the construciton with the other edges, which provides us with all the shifts transforming $P_0$ into its immediate neighbours. Now, we notice that, for this construction, we need all the shifts defined by the edges of $P_0$. They are enough as the shifts around the sides of $P$ are given by $\tau_1$ and $\tau_2^{\tau_1},\ldots,\tau_p^{\tau_1}$.

For the case when $q$ is odd, the situation is a bit more complex. In fact, we take this time the mid-points of the edges of $Q_0$, the basic polygon, into consideration. Now, we consider also the mid-points of all edges of polygons which share a vertex with $Q_0$. Now, fix a vertex $V_1$ of $Q_0$. We consider all the mid-point of the edges which have a vertex in common with $Q_0$. All such mid-points which are around $V_1$ constitute the **mid-point figure** around $V_1$, see figure 5, where a partial view is given.

Let us focus on this figure. $M_i$ and $M_{i+1}$ are consecutive mid-points of edges which share $V$. The mid-point line which joins $M_i$ and $M_{i+1}$ also meets the line $AP$ in $P$ and the line $BQ$ in $Q$. The line $AP$ is an edge of a copy $Q_b$ of $Q_0$ which shares $V$ with $Q_0$ and which is also determined by its other edge $VP$. Similarly, the line $BQ$ is also an edge of another copy $Q_a$ of $Q_0$ which shares $V$ with $Q_0$ and which is determined by its edege $VQ$. Now, the shift $\sigma_i$ along the line $M_i M_{i+1}$ which transforms $A$ into $M_{i+1}$ transforms $Q_b$ into $Q_a$. The opposite shift, along the same line, transforms $Q_a$ into $Q_b$ and, for instance, $B$ into $M_i$.

By rotation around $V$, we determine the other shifts, constructed from two consecutive mid-point edges around $V$. It is not difficult to note that by applying these shifts consecutively in turning twice around the vertex, we obtain all the copies of $Q_0$ which are around $V$. Now, one of these shifts, say $\tau$ transforms $Q_0$ in another neighbouring polygon $Q$. Note that all shifts, constructed around vertices in the above indicated way, but corresponding to $Q$, are obtained from those, say $t$, which are attached to $Q_0$ as $t^\tau$. Accordingly, the shifts attached to $Q_0$ by the above process generate all the shifts which leave the tiling $\{p,q\}$ invariant.

And so, we proved the following extension of lemma 4:

**Lemma 5** *For all positive integers $p$ and $q$ such that $\dfrac{1}{p} + \dfrac{1}{q} < \dfrac{1}{2}$, the shifts leaving the grid $\{p,q\}$ globally invariant are finitely generated. The number of generators is at most $p$ when $q$ is even, and at most $p.q$ when $q$ is odd.*

## 5 Commutation with shits and rotation invariance

First of all, we have to define what is rotation invariance and then, we shall prove that it is characterized by the commutation with shifts.

### 5.1 Rotation invariance

In the Euclidean plane, the definition of rotation invariant rules, a well known notion in cellular automata, can easily be defined.

Consider the case of von Neumann neighbourhood. It is not difficult to see that the rules of a cellular automaton can be represented as follows:

$(r) \quad s_N, s_E, s_S, s_W, s_c \rightarrow s'_c,$

$s_N$, $s_E$, $s_S$ and $s_W$ are the states of the neighbours of $c$ which are on the North, the East, the South and the West respectively. The state of $c$ itself is $s_c$ at the moment when the ruled is applied, and it becomes $s'_c$ after that, which is indicated by the arrow in formula $(r)$.

In the Euclidean case, a rotation invariant cellular automaton $A$ is **rotation invariant** if for all rules of $A$ written in the form of $(r)$, the rules obtained from $(r)$ by a circular permutation on the terms which are on the left-hand side of the arrow are also rules of $A$ and they all give the same new state $s'_c$ as in $(r)$.

It is not difficult to see that such a syntactic rule can easily be transported to the case of any grid $\{p,q\}$ of the hyperbolic plane.

If we transpose the definition of the Euclidean plane to the hyperbolic one, we can see that the notion of direction plays a key rôle. As mentionned in the introduction, there is no such notion on the hyperbolic plane. The tools introduced in [4] provide us with something which plays the rôle of the North pole in the hyperbolic plane. As the basic structure of a tiling $\{p,q\}$ of the hyperbolic plane is the existence of a generating tree, for each cell, the central one excepted, the direction to the father is a way to define a direction in a meaningful way. In the case of cellular automata in the Euclidean plane, the coordinates seems to be so an evident feature that almost nobody pays attention to that. However, if we want to **actually** implement cellular automaton for some simulation purpose, we are faced to the problem, even in this trivial case. And we can see that there is a price to pay, although the coordinate system seems to be for free. In a concrete implementation, cells have coordinates which are numbers, and numbers take some room which cannot be neglected. It could be answered that this is a hardware matter and that in a theoretical study, we may ignore this constraint. OK, let us take that granted. In this case, we can assume the same for the hyperbolic plane: fixing a central cell, the generating trees and from that the coordinates of any cell is a hardware feature.

In the next section, we shall go back to this question. We shall see that the question of direction can be, *theoretically* be handled in a pure *cellular automata* approach.

Remember that the neighbourhood of a cell $c$ is a part of a ball around $c$ which contains $c$ itself. We require that the neighbourhoods $N_c$ and $N_d$ of two cells $c$ and $d$ could be put into a one-to-one correspondance by a positive displacement $\delta$ from $N_c$ onto $N_d$ such that $\delta(c) = d$ and $\delta(d) = f(d)$, where $f(x)$ is the father of the cell $x$. As we shall consider the question of rotation invariance, we assume that the neighbourhood around a cell $c$ is a ball around $c$ of a fixed radius $k$ depending only on the cellular automaton. Now, as the father is know, we can number the neighbours of $c$ by associating 0 to the father and then, clockwise turning around the cell, by associating the next numbers to the next cells and then, going on in this way until we reach the last cell which is at the distance $k$ of $c$. This allows us to define the **format** of a rule as follows:

$$(R) \qquad \left( \{(\eta_i)\}_{i \,\in\, \{0..\alpha u_k\}}, \eta \right) \to \eta'$$

where $\eta_i$ is the state of the neighbour $i$ of $c$, $u_k$ is the number of cells in $\mathcal{F}_{pq}$ which are at distance $k-1$ from the root of $\mathcal{F}_{pq}$, and $\alpha$ is the number of such trees around the central cell. Note that, in particular, $\eta_0$ is the state of the father of $c$.

Now, it is easy to notice that a circular permutation on the numbers of the cells which are at distance 1 of $c$ can be extended into an isometry which is nothing else than a rotation around $c$, and conversely. Accordingly, we can give the following definition:

**Definition 1** *Consider a cellular automaton $A$ on a grid $\{p,q\}$ of the hyperbolic plane, and assume that the neighbourhood of any cell $c$ is a ball around $c$ of radius $k$, where $k$ is a constant. Say that $A$ is **rotation invariant** if and only if for any rule of its table which can be written in the form $(R)$, all the rules:*

$$(R') \quad \left(\{(\eta_\pi(i))\}_{i \,\in\, \{0..\alpha u_k\}}, \eta\right) \to \eta'$$

*where $\pi$ is a circular permutation on $\{0..\alpha u_k\}$, also belong to the table of $A$.*

### 5.2 Commutation with shifts

Consider a cellular automaton $A$ on the grid $\{p,q\}$ of the hyperbolic plane. Let us deonte by $\mathcal{C}$ the set of configurations on the grid. We define the **global function** $F_A$ from $\mathcal{C}$ into $\mathcal{C}$ as usual: if $x \in \mathcal{C}$, then for any cell $c$, we have $F_A(x)(c) = f(x(N_c), x(c))$, where $N_c$ is the set of the neighbours of $c$, listed as $\{c_i\}_{i \,\in\, \{0..\alpha u_k\}}$, according to the numbering which we have above defined, and $f$ is the table of the rules of $A$.

**Definition 2** *Let $A$ be a cellular automaton on the grid $\{p,q\}$ of the hyperbolic plane. Let $F_A$ denote its global transition function. Then $A$ is said to **commute with the shifts** if and only if $F_A \circ \sigma = \sigma \circ F_A$ for all shifts $\sigma$ leaving the grid $\{p,q\}$ globally invariant.*

The main result of this section is:

**Theorem 2** *A cellular automaton on the grid $\{p,q\}$ of the hyperbolic plane commutes with the shifts if and only if it is rotation invariant.*

Before proving the theorem, let us remark that most cellular automata which are devised for various theoretical computations are rotation invariant. This is the case for many of them in the Euclidean plane. It is also the case of several of them, among the few ones devised in the hyperbolic plane or in the $3D$ space.

Let us go back to the definition of the commutation of $F_A$ with a shift. This means that: $F_A(\sigma_({x})) = \sigma(F_A(x))$. Let $d = \sigma(c)$, where $c$ is a cell. Then $F_A(\sigma(x))(d) = f(\sigma(x(N_c)), s_c)$, where $f$ is the table of $A$, as $\sigma$ gives in $d$ the state which we have in $c$. Now, $\sigma(x(N_c))$ clearly transports the states of the cells in $N_c$ onto a set of states on a rotated image of $N_d$ with respect to the father of $d$. And, a priori, the father of $d$ is not the image of the father of $c$ under $\sigma$. In the next sub-section, we shall see that indeed, the shifts need not commute with the operation which, to a cell, assigns its father.

Accordingly, if the cellular automaton commutes with the shifts, it is invariant under this rotation, and conversely. Now, we know that all these rotations are generated by shifts, as it easily follows from the proof of lemma 5. Consequently, this gives us the result. ∎

### 5.3 Rotation invariant cellular automata

In this section, we shall first see that a cellular automaton on a grid $\{p,q\}$ need not commute with shifts. Then, we shall prove the following result:

**Theorem 3** *For any cellular automaton $A$ on the pentagrid or the ternary heptagrid, there is a cellular automaton $B$ and a projection $\xi$ of the states of $B$ on state of $A$ such that $B$ is rotation-invariant and, for any cell $c$, $A(c) = \xi(B(c))$. There is also another cellular automaton $C$ with a projection $\chi$ of its states on those of $A$ satisfying $A(c) = \chi(C(c))$ and which is not rotation invariant.*

The proof of the theorem is obtained by constructing a **product** automaton with a cellular automaton which we shall define. Then, from this product, we shall construct a set of rules which is not rotation invariant and another one which is so.

The special factor of this product is a cellular automaton which propagates the tree structure inside the grid, here the pentagrid or the ternary heptagrid.

For this purpose, we assign an **extended status** to each cell which is an extension of the notion of status of this cell as a node of the Fibonacci tree where it stands. Remember that a node is **black** if it has two sons and that it is **white** if it has three sons. Black and white defines the **status** of the node, see [4]. Now, we define the **extended status** as follows, indicating them by **symbols** at the same time. First, we proceed with black nodes and then with white ones.

> $Bb$, $Bw$ : black node, black, white father respectively; in figure 6, below, they are represented by the colours dark and light blue, respectively.
> 
> $Wwm$, $Wwr$ : white node, white father, middle, right-hand son, respectively; in figure 6, they are represented by the colours yellow and green, respectively.
> 
> $Wb$ : white node, black father, represented in orange in figure 6.

For each node, its immediate neighbours are given by the following tables, first listing the father $f$ of a cell $c$ and then, clockwise turing around $c$, its neighbours $n_1, \ldots, n_{\alpha-1}$, with $\alpha \in \{5, 7\}$.

We can see that black nodes are always identified by the pattern $Bb$, $Wb$, $Bw$ in their immediate neighbourhood, while white nodes are identified by the pattern $Bw$, $Wwm$, $Wwr$.

Now, the extended status can always be inferred from such a neighbourhood. In nodes of extended status $Bb$ and $Bw$, the identification comes from the neighbour $n_1$ : it is white for $Bb$-nodes but $Wwm$ nether occurs. For white nodes, the distinction between the extended status $Wwm$ and the others comes from the neighbour $n_4$: it is $Bw$ for $Wwm$ nodes and $Bb$ for the others. Between $Wmr$ and $Wb$ nodes, the difference comes from the father, of course.

Now, the rows of these tables can easily be transformed into conservation rules: a row $c, f, n_1, \ldots, n_{\alpha-1}$ induces the rule $f, n_1, \ldots, n_{\alpha-1}, c \to c$.

It remains to see that we can define **propagation rules** for a cellular automaton. Indeed, the initial configuration would assign a special state to the central cell and the quiescent state to all the other cells. Then, the propagation rules would define the extended status of the neighbouring cells, and defining the extended status of all cells, ring by ring, where a ring is a set of cells at the same distance from the central cell.

| $\nu$ | $f$ | $n_1$ | $n_2$ | $n_3$ | $n_4$ |
|---|---|---|---|---|---|
| $Bb$: | $Bb$ | $Wmr$ | $Bb$ | $Wb$ | $Bw$ |
| | $Bw$ | $Wb$ | $Bb$ | $Wb$ | $Bw$ |
| | $Bw$ | $Wmr$ | $Bb$ | $Wb$ | $Bw$ |
| $Bw$: | $Wwm$ | $Bw$ | $Bb$ | $Wb$ | $Bw$ |
| | $Wwr$ | $Wwm$ | $Bb$ | $Wb$ | $Bw$ |
| | $Wb$ | $Bb$ | $Bb$ | $Wb$ | $Bw$ |
| $Wwm$: | $Wwm$ | $Bw$ | $Wwm$ | $Wwr$ | $Bw$ |
| | $Wwr$ | $Bw$ | $Wwm$ | $Wwr$ | $Bw$ |
| | $Wb$ | $Bw$ | $Wwm$ | $Wwr$ | $Bw$ |
| $Wwr$: | $Wwm$ | $Bw$ | $Wwm$ | $Wwr$ | $Bb$ |
| | $Wwr$ | $Bw$ | $Wwm$ | $Wwr$ | $Bb$ |
| | $Wb$ | $Bw$ | $Wwm$ | $Wwr$ | $Bb$ |
| $Wb$: | $Bb$ | $Bw$ | $Wwm$ | $Wwr$ | $Bb$ |
| | Bw | $Bw$ | $Wwm$ | $Wwr$ | $Bb$ |

**Table 2** *Rules for the conservation of the structure of the Fibonacci tree, case of the pentagrid.*

We give the propagation rules for such an automaton in the case of the pentagrid and we leave the case of the ternary heptagrid as an exercise for the reader.

Now, we are in the position to prove theorem 3.

Consider the automaton $P$ whose table is defined by the rules of figure 6 and table 2 in the case of the pentagrid. In the case of the ternary heptagrid, the propagation rules are adapted from figure 6 and also taken from table 3.

Let $A$ a cellular automaton. We first define the product $A \times P$ by the states $(\eta, \pi)$, where $\eta$ runs over the states of $A$ and $\pi$ over those of $P$. We shall also say that $\eta$ is the $A$-state of $(\eta, \pi)$ and that $\pi$ is its $P$-states.

Before going further, let us note that the function which associates its father to a cell does not necessarily commute with shifts.

This can easily be seen on figure 4. Consider its left-hand side picture, the case of the pentagrid. Imagine that $\Pi_0$ is a black node whose father is $\Pi_1$. Then the shift $ED$, which transforms $E$ into $D$ along the line passing throuhg these points transforms $\Pi_0$ into its black son $\Pi_5$. Now, the same shift does not transform $\Pi_1$ into $\Pi_0$, but in the reflection of $\Pi_1$ in the line $BD$. On another hand, the shift $BD$ transforms $\Pi_1$ into $\Pi_0$ and $\Pi_0$ into $P_4$ whose father is indeed $\Pi_0$. The same figure shows that for each kind of node and each kind of son there is a shift which maps the father onto the father in this situation and a shift which does not.

| $\nu$ | $f$ | $n_1$ | $n_2$ | $n_3$ | $n_4$ | $n_5$ | $n_6$ |
|---|---|---|---|---|---|---|---|
| $Bb$: | $Bb$ | $Wwr$ | $Wwr$ | $Bb$ | $Wb$ | $Bw$ | $Wb$ |
|  | $Bw$ | $Wb$ | $Wwr$ | $Bb$ | $Wb$ | $Bw$ | $Wb$ |
|  | $Bw$ | $Wwr$ | $Wwr$ | $Bb$ | $Wb$ | $Bw$ | $Wb$ |
| $Bw$: | $Wwm$ | $Bw$ | $Wb$ | $Bb$ | $Wb$ | $Bw$ | $Wwm$ |
|  | $Wwr$ | $Wwm$ | $Wwr$ | $Bb$ | $Wb$ | $Bw$ | $Wwm$ |
|  | $Wb$ | $Bb$ | $Wb$ | $Bb$ | $Wb$ | $Bw$ | $Wwm$ |
| $Wwm$: | $Wwm$ | $Bw$ | $Bw$ | $Wwm$ | $Wwr$ | $Bw$ | $Wwr$ |
|  | $Wwr$ | $Bw$ | $Bw$ | $Wwm$ | $Wwr$ | $Bw$ | $Wmr$ |
|  | $Wb$ | $Bw$ | $Bw$ | $Wwm$ | $Wwr$ | $Bw$ | $Wmr$ |
| $Wwr$: | $Wwm$ | $Wwm$ | $Bw$ | $Wwm$ | $Wwr$ | $Bb$ | $Bw$ |
|  | $Wwr$ | $Wwm$ | $Bw$ | $Wwm$ | $Wwr$ | $Bb$ | $Bb$ |
|  | $Wb$ | $Wwm$ | $Bw$ | $Wwm$ | $Wwr$ | $Bb$ | $Bb$ |
| $Wb$: | $Bb$ | $Bb$ | $Bw$ | $Wwm$ | $Wwr$ | $Bb$ | $Bw$ |
|  | $Bw$ | $Bb$ | $Bw$ | $Wwm$ | $Wwr$ | $Bb$ | $Bw$ |

**Table 3** *Rules for the conservation of the structure of the Fibonacci tree, case of the ternary heptagrid.*

This allows us to prove the theorem. First, we notice that we can consider cells at a time when their $P$-state is stable. Then, we note that the rules of $A \times B$ are of the form:

$$(R_1) \quad \{(\eta_i, \pi_i)\}_{i \in \{0..\alpha-1\}}, (\eta, \pi) \to (\eta', \pi)$$

From the table 2 and 3, it is clear that rotating a rule does not entail contradictions with already established rules: the distinction between the actual father and the *rotated* one is always clear.

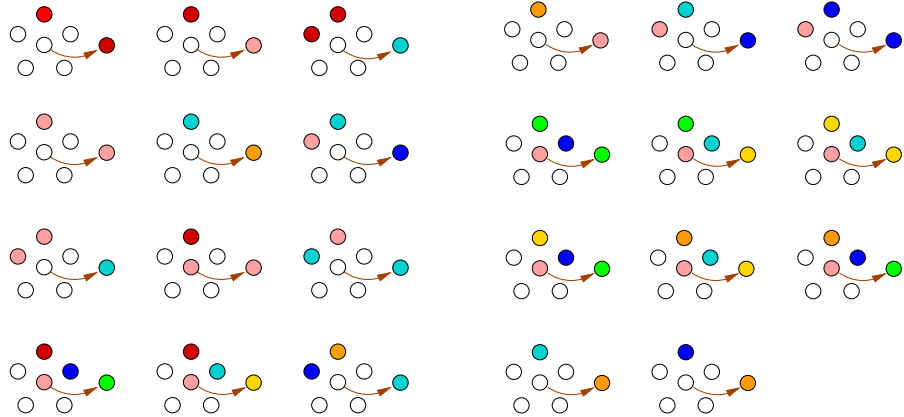

**Figure 6** *Rules for the propagation of the Fibonacci tree structure in the case of the pentagrid.*

Accordingly, we can decide, either to introduce all the following rotated rules:

$(R')$ $\{(\eta_i, \pi_{\sigma(i)})\}_{i \,\in\, \{0..\alpha-1\}}, (\eta, \pi) \to (\eta', 0_P)$,

where $0_P$ is the quiescent state of $P$ and $\sigma$ does perform a rotation, or all the following ones:

$(R')$ $\{(\eta_i, \pi_{\sigma(i)})\}_{i \,\in\, \{0..\alpha-1\}}, (\eta, \pi) \to (\eta', \pi)$.

In the first case, the new automaton is not rotation invariant. In the second case, it is rotation invariant. ∎

As a matter of case, for the cellular automaton $P$ itself, the rules given by figure 6 are rotation invariant, while those given by tables 2 and 3 are not. The just produced argument for the proof of theorem 3 allows us to extend the rules of tables 2 and 3 either to rotation invariant ones or to non rotation invariant ones.

## 6 Proving Hedlund's theorem

Now, the proof of the theorem goes as it does classically.

From lemmas 1 and 2, we know that cellular automata on grids $\{p, q\}$ are continuous on the space of configurations. From lemma 3, we know that they commute with any shift if and only if they are rotation invariant.

For the converse, we consider a mapping $F$ on the space of configurations. We assume that it is continuous with respect to the topology defined in section 3 and that it commutes with the shifts. Then, again, the standard argument applies. The compacity of the space with respect to the topolgy allows to consider the distance between two sets $\{x \mid F(x)(c) = p\}$ for different states $p$, as the configurations are defined on $Q^{\mathcal{F}_{pq}}$, $Q$ being called the set of states which we assume to be finite, $c$ being a fixed cell. This minimal distance is positive and it allows to define a ball $B_n$ for some $n$ such that the value of $F(x)$ at $c$ depends only on the values of $x$ on the ball $B_n$ around $c$.

Next, as in the classical proofs, we transport this property to any cell thanks to the commutation property of $F$ with the shifts. ∎

And so, we proved theorem 1. From this, we immediately get, as classically:

**Theorem 4** *A rotation invariant cellular automaton on a grid $\{p,q\}$ of the hyperbolic plane is reversible if and only if it is bijective.*

## 7 Conclusion

The question arises whether other classical theorems on cellular automata are also true for hyperbolic cellular automata. As an example, we can take the theorems of Moore and Myhill, see [8,9], characterizing surjective global transition functions as injective global transition funcions restricted to finite configurations. In fact, it seems difficult to adapt the classical proof in a straightforward way.

The reason is that the classical argument relies on the fact that the surface of a big square in a square tiling of the Euclidean plane becomes negligeable with respect to its all area when the size of the square tends to infinity. In the hyperbolic plane, this is no more true for a closed ball: the number of tiles on the border is more than the half of the total of number of all the tiles in the ball.

And so, there is still some work ahead: either to find another argument, or to find that Moore's or Myhill's theorem is no more true in the hyperbolic space.

Another example is the theorem about whether the reversibility of cellular automata in the hyperbolic plane is undecidable as it is in the case for the Euclidean plane, see [3].

Accordingly, there is still much work to do in these directions.